\documentclass[apjl,twocolumn]{aastex6}
\usepackage{graphicx,color,ulem}
\usepackage{amsmath}

\begin{document}
\newcommand{\kenta}[1]{\textcolor{blue}{#1}}
\newcommand{\notekh}[1]{\textcolor{magenta}{#1}}
\newcommand{\tsvi}[1]{\textcolor{red}{#1}}
\newcommand{\notetp}[1]{\textcolor{cyan}{#1}}
\newcommand{\deltp}[1]{{\color{red}  \sout {#1}}}

\title{Are the observed black hole mergers spins consistent with field binary  progenitors? }
\author{Kenta Hotekezaka\altaffilmark{1} and Tsvi Piran\altaffilmark{2}
            }
            \altaffiltext{1}
	{Center for Computational Astrophysics, Flatiron Institute, New York, 10010, NY,  USA}
	\altaffiltext{2}
	{Racah Institute of Physics, The Hebrew University of
		Jerusalem, Jerusalem 91904, Israel}
	\email{E-mail:khotokezaka@flatironinstitute.org, tsvi.piran@mail.huji.ac.il}


\begin{abstract}
One of the puzzles in the recent observations of gravitational waves from binary black hole mergers is the observed low (projected)
spins of the progenitor black holes. 
In two of the four events, GW150914,  and the recent GW170104,  the observed spins  are most likely negative (but consistent with zero). In the third case LVT151012 it is practically zero and only in the  forth case, GW151226,  the spin is positive but low.   These observations are puzzling within the field binary scenario in which positive  higher spins are expected. Considering the most favorable Wolfe Rayet  (WR) progenitors
we estimate the  expected spin distribution for different  evolution scenarios and compare it to the observations. 
With typical parameters one expects a significant fraction ($\ge 25\%$) of the mergers  to have high effective spin values. 
However due to  uncertainties in the  outcome of the common envelope phase (typical separation and whether the stars are rotating or not) and   in the late stages of massive star evolution  (the strength of the winds) one cannot rule our scenarios in which the expected spins are low. While observations of high effective spin events will support this scenario, further observations of negative spin events would rule it out.

\end{abstract}



\section{Introduction}\label{sec:Intro}
Gravitational-wave astronomy begun  in Sept 14th 2015 with LIGO's discovery \citep{abbott2016PhRvL} of  GW150914, a binary black hole (BBH) merger.
An additional BBH merger, GW151226, as well as a merger candidate, LVT151012 were discovered in LIGO's O1 run. A forth event, GW170104 \citep{Abbott2017PRL}, was discovered in the O2 run that begun in the late fall of 2016 and still continues now. All BBH mergers discovered so far involved rather massive BHs with the lightest one (observed in GW151226) is of  7.5$m_\odot$. 

Among the remarkable features of all four events are the relatively low values of $\chi_{\rm eff}$, the mass-wighted
average of the dimensionless spin components $\chi\equiv a/m$, projected along the orbital angular momentum,   of the  individual BHs before they merged. 
While in none of the cases $\chi_{\rm eff}$ is large, in two cases the best fit values are negative (but the error bars don't exclude zero), in one case it is practically zero and only in the forth case this value is positive but  small. These values are best fitted by  a low-isotropic spin distribution (\citealt{farr2017arXiv} and see also \citealt{vitale2017}) and are at some ``tension"   with
the expectations  from  field binary evolution scenarios that suggest that the individual spins should be aligned with the orbital angular momentum axis and at least in a significant fraction of the events the spins should be large \citep{zaldarriaga2017,HotokezakaPiran2017ApJ}. 
We compare here the observed distribution with the one expected in the most favorable evolutionary scenarios involving WR stars and discuss whether the observations disfavor  field binary evolution models \citep{belczynski2016Nature,belczynski2017,stevenson2017NatCo,postnov2017arXiv}  and  support  capture
models  \citep{rodriguez2016ApJb,oleary2016ApJ,
antonini2016ApJ,bartos2017ApJ,sasaki2016PhRvL,bird2016PhRvL,blinnikov2016JCAP,kashlinsky2016ApJ} in which the spins are expected to be randomly oriented. 

The essence of the argument concerning that suggests a  large $\chi_{\rm eff}$ in field binary BH is the following:
(i) To merge within a Hubble time, $t_{\rm H}$,   the initial semi-major axis of the BBH at the moment of the formation of the second BH, $a$,  should not be too large. 
(ii)  With a relatively small $a$   the stars feel a significant tidal force and  their
spin tends to be synchronized with the orbital motion.  
(iii) If synchronized, the short  orbital period implies that the progenitor's spin $\chi_{*} \equiv S c/Gm^2 $ (where $m$ is the BH's mass, $S$ its spin angular momentum, $G$ is the gravitational constant, and $c$ is the speed of light) is large:  $\chi \ge \chi_{*}(t_{H})$, where $\chi_{*}(t_{H})$ is the  spin parameter of a  star in a  binary that will merge in a Hubble time. Therefore,   $\chi_{*}(t_{\rm H})/2\lesssim \chi_{\rm eff}\lesssim 1/2$ if only the secondary  has been synchronized. $\chi_*(t_{\rm H})\lesssim \chi_{\rm eff} \lesssim 1$ if both progenitors have been synchronized. 

We discuss first, in \S \ref{sec:obs}, the gravitational wave observations as well as observations of  galactic X-ray binaries containing BHs. In \S \ref{sec:theory},
following this chain of arguments, we  express the initial semi-major axis, $a$, in terms of the the merger time, $t_c$,
and we estimate  $\chi_*(t_{c})$.  We  express   the dimensionless BH spin $\chi$ in terms of the progenitor's parameters.      In \S \ref{sec:results}, using these estimates we calculate the expected spin distribution in different scenarios and compare it to the gravitational-wave observations. 

\section{Observations} 
\label{sec:obs}

{\it Binary BH Mergers:} Some basic observed properties of the BBH merger events are summarized in Table I. The most interesting ones for our purpose are the BHs' masses, their  $\chi_{\rm eff}$ values and the final BH spin.
This latter quantity is  defined as: 
 \begin{equation} 
 \chi_{\rm eff} \equiv \frac{ m_1 \chi_1 + m_2 \chi_2 }{m_{\rm tot}}  \ \ \ \ {\rm where } \ \ \ \ \chi_{1,2} \equiv \frac{c\vec S_{1,2} \cdot \hat L } { G m_{1,2}^2}, 
\end{equation}
and $m_{tot}=m_1+m_2$ and $\hat L$ is a unit vector in the direction of the system's orbital angular momentum $\vec L$.
  The limits on $\chi_{\rm eff}$ are obtained from the observations of the gravitational wave signals before the merger. The lack of extended ringdown phases also puts   limits on the spins of the final BHs, $a_f$. The fact that those are  of order $0.6-0.7$ and not close to unity is an independent evidence that the initial aligned spins of the BHs were not close to unity.
Had the initial aligned spins been large, the final spin of the merged BHs would have been very close to unity and would have had a long ringdown phase. Thus the final spin and the initial spins estimates are consistent.  Indeed, the final spin is slightly larger ($0.74^{+0.06}_{-0.06}$) for GW151226, the only case for which the nominal value of $\chi_{\rm eff} =0.21^{+0.20}_{-0.10}$ is positive.

\begin{figure}[h]
  \begin{center}
    \includegraphics[width=0.8\linewidth]{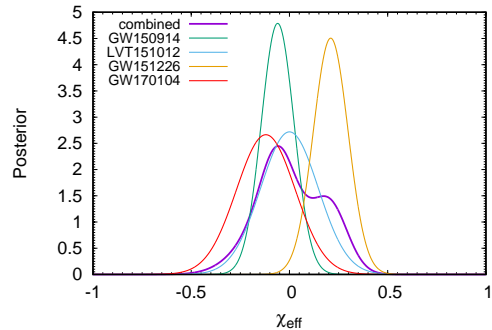}
    \caption{The distribution of the observed spins. We have approximated each observed distribution as a Gaussian whose mean value and $90\%$ confidence interval are the same to the values shown in 
     \cite{abbott2016PhRvX,Abbott2017PRL} (see \citealt{farr2017arXiv}).
     Also shown is a combined  distribution of the four Gaussians.  } 
    \label{fig:obsspins}
  \end{center}
\end{figure}

\begin{table*}
\begin{center}
\label{tab:LIGO}
{\begin{tabular}{lcccccc}
\hline \hline
Event & $m_1$           & $m_2$                 & $m_{\rm tot}$ & $\chi_{\rm eff}$ & $a_f$  \\
          & $[m_{\odot}]$ &   $[m_{\odot}]$    &$[m_{\odot}]$  &            &          \\  \hline  
GW150914 & $36.2^{+5.2}_{-3.8}$ & $29.1^{+3.7}_{-4.4}$ & $65.3^{+4.1}_{-3.4}$ & $-0.06^{+0.14}_{-0.14}$ &$0.68^{+0.05}_{-0.06}$ \\
GW151226 & $14.2^{+8.3}_{-3.7}$ & $7.5^{+2.3}_{-2.3}$ & $21.8^{+5.9}_{-1.7}$   & $0.21^{+0.20}_{-0.10}$ &$0.74^{+0.06}_{-0.06}$ \\
LVT151012 & $23^{+18}_{-6}$ & $13^{+4}_{-5}$ & $37^{+13}_{-4}$                         & $0.0^{+0.3}_{-0.2}$ &$0.66^{+0.09}_{-0.10}$ \\
GW170104 & $31.2^{+8.4}_{-6.0}$ & $19.4^{+5.3}_{-5.9}$ & $50.7^{+5.9}_{-5.0}$  & $-0.12^{+0.21}_{-0.30}$ & $0.64^{+0.09}_{-0.20}$ \\
\hline \hline 
\end{tabular}}
\end{center}
\caption{Parameters of the BBH mergers detected during LIGO's O1 and O2 runs. The parameters are median values with 90\% confidence intervals. 
The values are taken from \cite{abbott2016PhRvX,Abbott2017PRL}.}
\end{table*}
Fig. \ref{fig:obsspins}  describes the observed $\chi_{\rm eff}$ distribution in terms of  the corresponding four  Gaussians describing approximately the $\chi_{\rm eff}$ posterior distributions of the observed events and the resulting combined spin distribution for the whole sample.   

{\it Galactic  BHs in X-ray binaries:}  Observations of X-ray binaries involving BHs, albeit smaller mass ones, can also shed some light on the problem at hand.  In particular observations of two such systems that include massive ($>10 m_\odot$) BHs, Cyg X-1 and GRS 1915+105,   provide a good evidence that these massive BHs formed in situ,  in a direct implosion and without a kick \citep{mirabel2016arxiv}.  For example, Cyg X-1 moves at $9 \pm 2$ km/s relative to the stellar association Cygnus OB3, indicating that it could have lost at most $1 \pm 0.3 m_\odot$ at formation. 
Furthermore, the minuscule eccentricity of Cyg X-1, $0.018 \pm 0.0003$,   \citep{orosz2011ApJ}  suggests that the orbit has been circularized during the binary evolution and the collapse didn't give the system a significant kick that disturbed the circular orbit.  {In addition, \cite{mandel2016MNRASb} shows that large natal kicks, $>80$ km/s, are not required to  explain the observed positions of low-mass X-ray binaries.  }

Estimates of the spins of the relevant BHs  \citep{McClintock2014SSRv} suggest that in these two systems  $a/m>0.95$. Three other BHs, LMC X-1, M33 X-7, and 4U 1543-47, whose masses are larger than $9 m_\odot$, have $\chi>0.8$.  Only one BH with a mass $> 9 m_\odot$, XTE J1550-564 has a significantly lower values ($\chi = 0.34^{+.20}_{-.28}$).  It is important to note that these large spins must be obtained at birth as accretion cannot spin up a massive BH to such a high spin value.

\section{Merger Time, Orbital Separation and Synchronization. }
\label{sec:theory}

Assuming circular orbits the merger time, due to gravitational radiation driven orbital decay, 
is: 
\begin{equation}
t_c  
\approx 10~ {\rm Gyr}~ \left(\frac{2 q^2}{1+q}\right)\left(\frac{a}{44R_{\odot}}\right)^4
\left(\frac{m_2}{30M_{\odot}}\right)^{-3}\ , 
\end{equation}
where $q\equiv m_2/m_1$.
Note that  we assume circular orbits here and elsewhere. This simplifying assumption is based on the expectation that the orbit will be circularized during the binary evolution and that it won't be affected by the collapse on the secondary. It is supported by the observations of binaries containing massive BHs, reported earlier (see \S \ref{sec:obs}).

Tidal forces exerted by the primary will tend to synchronize the secondary.  
If fully synchronized the final stellar spin would equal: 
\begin{eqnarray}
\label{Eq:spin}
\chi_2 &\approx &
0.5~
q^{1/4}\left(\frac{1+q}{2} \right)^{1/8} \left(\frac{\epsilon}{0.075}\right)   \left(\frac{R_2}{2R_{\odot}}\right)^{2} \nonumber \\
&&\left(\frac{m_2}{30M_{\odot}}\right)^{-13/8} \left(\frac{t_c}{1\,{\rm Gyr}}\right)^{-3/8}  \ ,
\end{eqnarray}
where $\epsilon$ characterizes the star's moment of inertia  $I_2 \equiv \epsilon m_2 R_2^2$. 
The progenitor's spin,  $\chi$, increases with the progenitors size and decreases when $t_c$ increases.  Thus, a compact progenitor star that formed at a high redshift produces  a low spin BH while a large progenitor  formed recently collapses to a large spin BH (see \citealt{kushnir2016MNRAS}). 

The synchronization process takes place over $t_{\rm syn}$:
\begin{eqnarray}
t_{\rm syn} &\approx&  20~{\rm Myr}~
\left(\frac{\epsilon}{0.075}\right) \left(\frac{E_{2}}{10^{-5}}\right)^{-1}
\left(\frac{R}{2 R_{\odot}}\right)^{-7}  
\nonumber \\ &&
\left(\frac{(1+q)^{31/24}}{q^{33/8}}\right)  
\left(\frac{m_2}{30M_{\odot}}\right)^{47/8}
\left(\frac{t_c}{1 {\rm Gyr}}\right)^{17/8} \ ,
\label{eq:Zahn}
\end{eqnarray}
where $E_2$, is a dimensionless quantity  introduced by \cite{zahn1975A&A}
characterizing the inner structure of the star.  
$E_2$ is  $\sim 10^{-7}$--$10^{-4}$ for 
massive main sequence stars and Wolf-Rayet (WR) stars \citep{zahn1975A&A,kushnir2017MNRAS}. 
The characteristic values used in Eq. (\ref{eq:Zahn}) correspond to a WR star. 
For WR stars, $t_{\rm synWR} $ can be expressed, following \citep{kushnir2016MNRAS}:
\begin{eqnarray}
t_{\rm synWR} \approx 10~{\rm Myr}~q^{-1/8}\left(\frac{1+q}{2q}\right)^{31/24}\left(\frac{t_c}{1~{\rm Gyr}}\right)^{17/8}. 
\label{eq:synWR}
\end{eqnarray}

Because of their short stellar lifetime, WR stars are not necessarily synchronized in binary systems even with $t_c$ of a few hundreds Myr. Therefore
the final stellar spin depends   on (i) $\chi_i$, the spins of the stars at the beginning of the WR phase (ii)  on the ratio of   $t_{\rm syn}$ and the lifetime of the WR star, $t_{\rm WR}$ and (iii) on the  angular momentum 
loss timescale during the WR phase, $ t_{\rm wind}$ \cite[see][]{kushnir2016MNRAS,HotokezakaPiran2017ApJ}.
With these parameters, we solve the following equation to 
obtain the stellar spin parameter at the end of the WR phase \citep{kushnir2016MNRAS}:
\begin{eqnarray}
\dot{\chi}_* = \frac{\chi_{\rm syn}}{t_{\rm syn}}\left(1-\frac{\chi_*}{\chi_{\rm syn}} \right)^{8/3} - \frac{\chi_*}{t_{\rm wind}},
\end{eqnarray}
where $\chi_{\rm syn}$ is the stellar spin parameter in the synchronized state.

\section{Collapse  and  the BH Spin}
\label{sec:collapse}
One can expect that, unless there is too much angular momentum (that is for $\chi_* \le 1$),   
the collapsing star implodes and the BH that forms  swallows all the collapsing stellar mass\footnote{The original stellar mass could be larger but this lost in an earlier phase due to winds \citep{mirabel2016arxiv}.}. If $\chi _*> 1$  a fraction of the matter will be ejected carrying  the excess angular momentum and  leading to a BH with $\chi \le 1$  \citep{stark1985PhRvL,oconnor2011ApJ,sekiguchi2011ApJ}. Thus we expect that 
\begin{equation}
\chi_{\rm BH} \approx  
\begin{cases}
 1  & \mbox{if } ~\chi_* \ge 1 ,  \\
\chi_*  &  \mbox{if }  ~\chi_* < 1 . 
\end{cases}  
\end{equation}

One may wonder if there are caveats to this conclusion. First, is it possible that matter
is ejected during the collapse to a BH even if 
 $\chi_* < 1$? This will, of course,
change the relation between the progenitor's spin and the BH's spin. Second is mass ejected isotropically? 
If not the BH will receive a kick and the BBH will be put into an elliptical orbit (that will merge faster). The kick may
also change the resulting BH spin. Since the initial spin is in the direction of the
orbital angular momentum the kick may reduce the spin component along this direction. Clearly these issues can be addressed by a
detailed numerical study of collapse to a BH. However, as discussed in \S \ref{sec:obs} observations of   binaries containing massive  ($>10 m_\odot$) BHs, Cyg X-1 and GRS 1915+105 provide a good evidence that massive BHs form in situ  in a direct implosion and without a kick  \citep{mirabel2016arxiv}.  
Estimates of spins of accreting massive BHs give an independent support to this conclusion. 

\begin{figure*}
  \begin{center}
        \includegraphics[width=6cm]{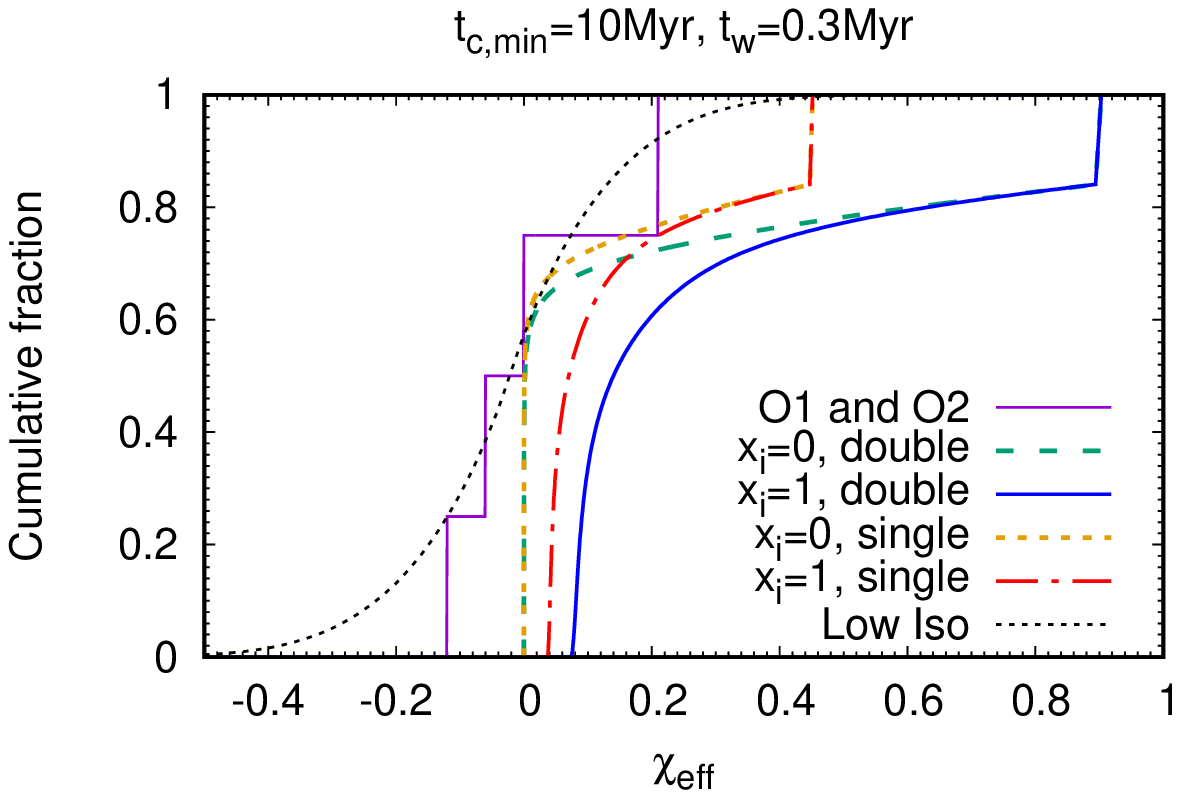}
         \includegraphics[width=6cm]{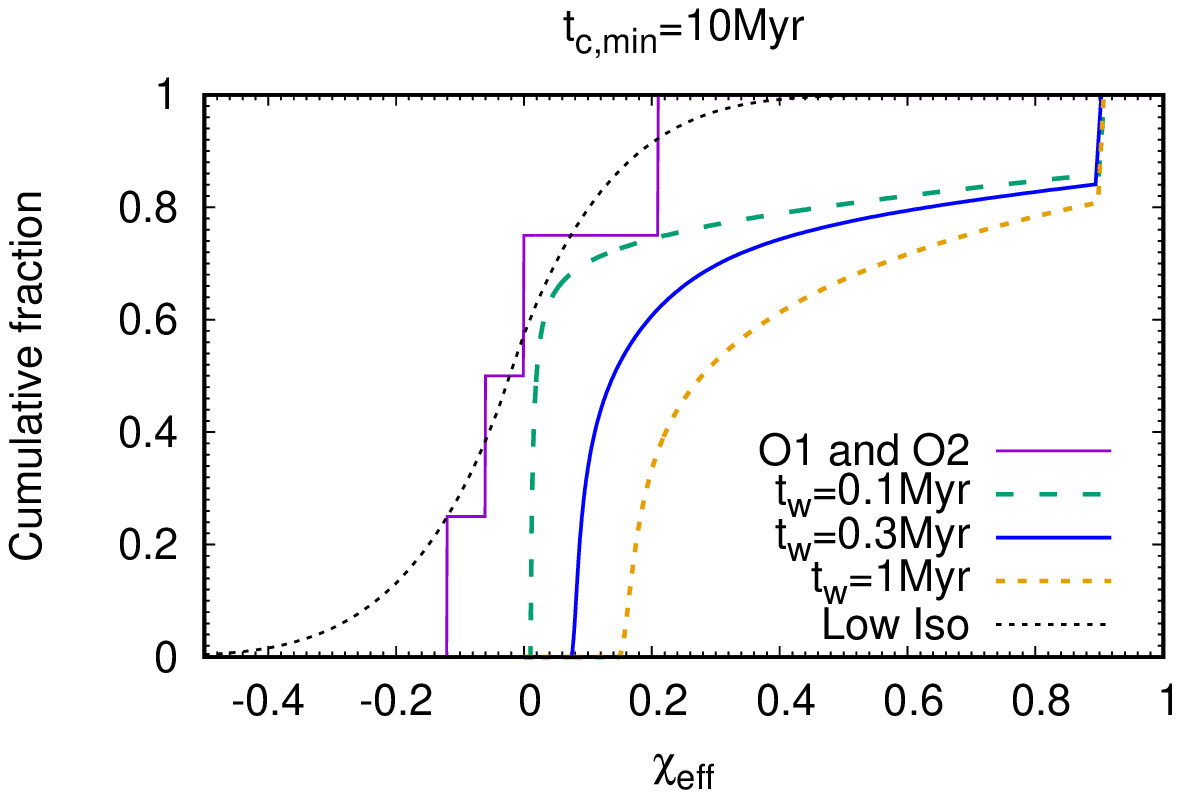}\\
                 \includegraphics[width=6cm]{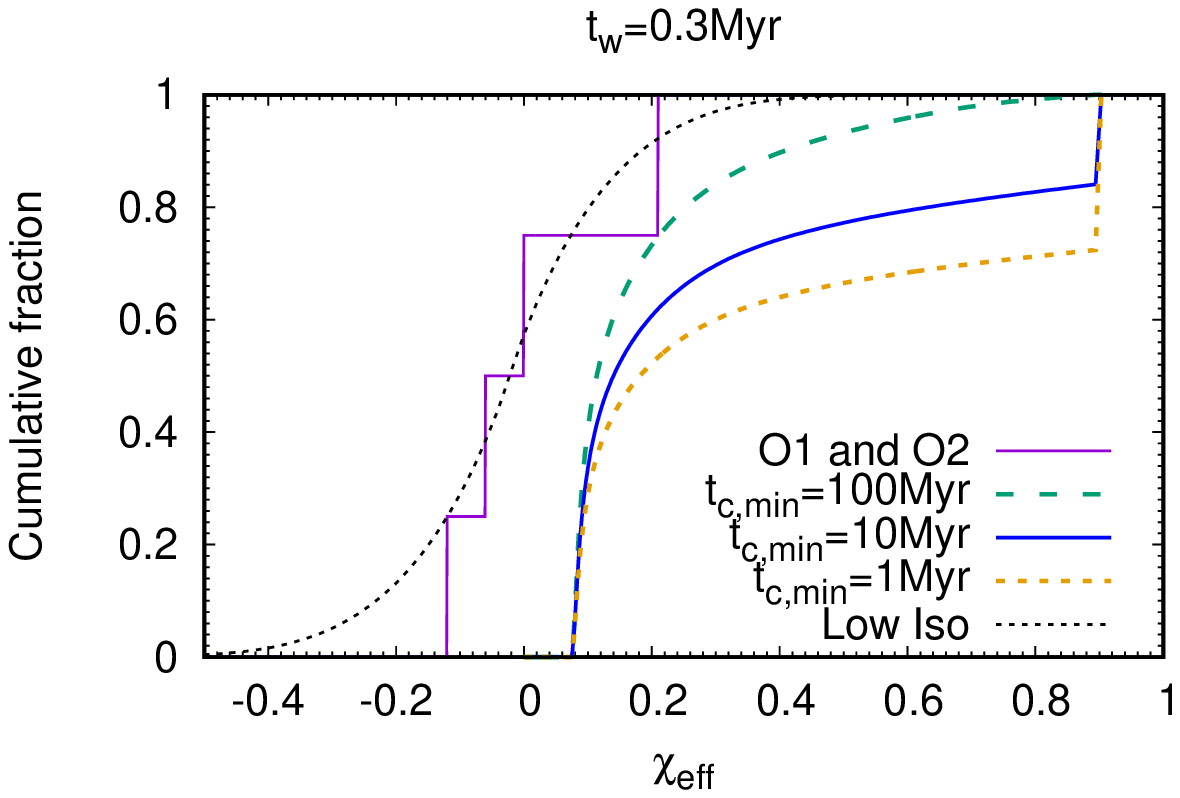}
         \includegraphics[width=6cm]{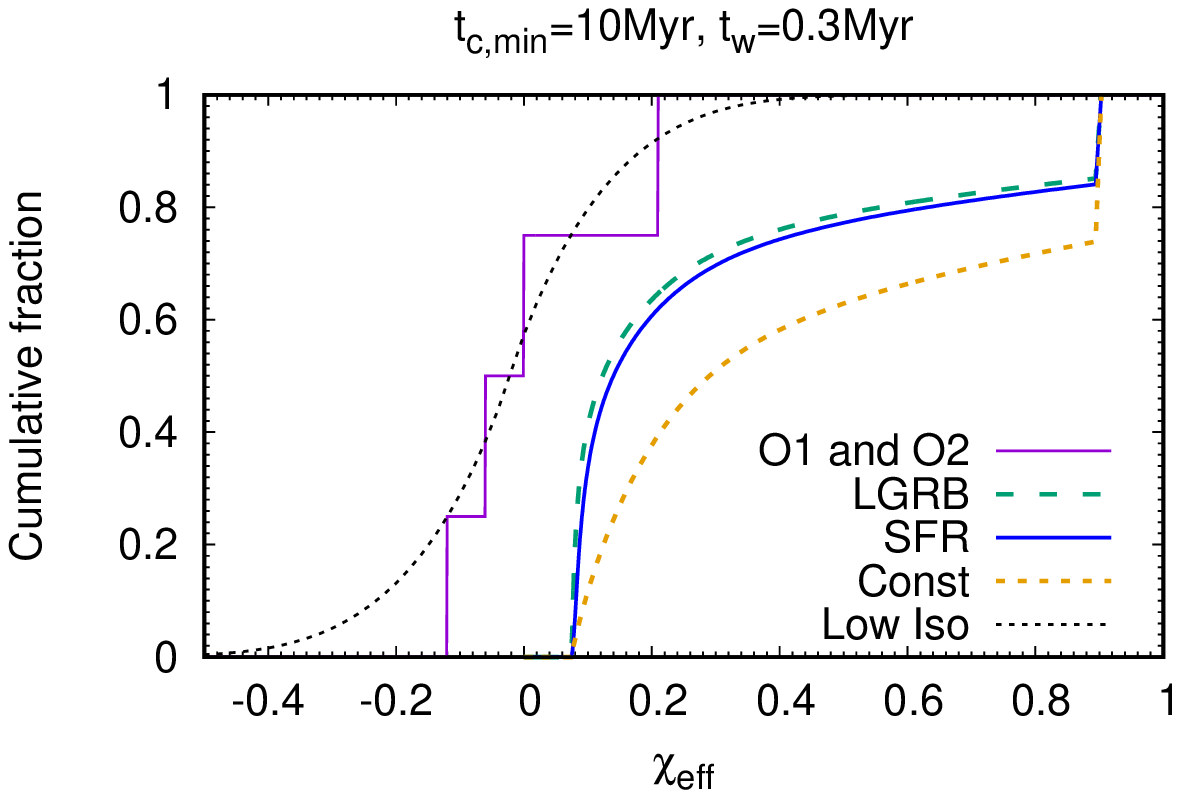}
         \includegraphics[width=6cm]{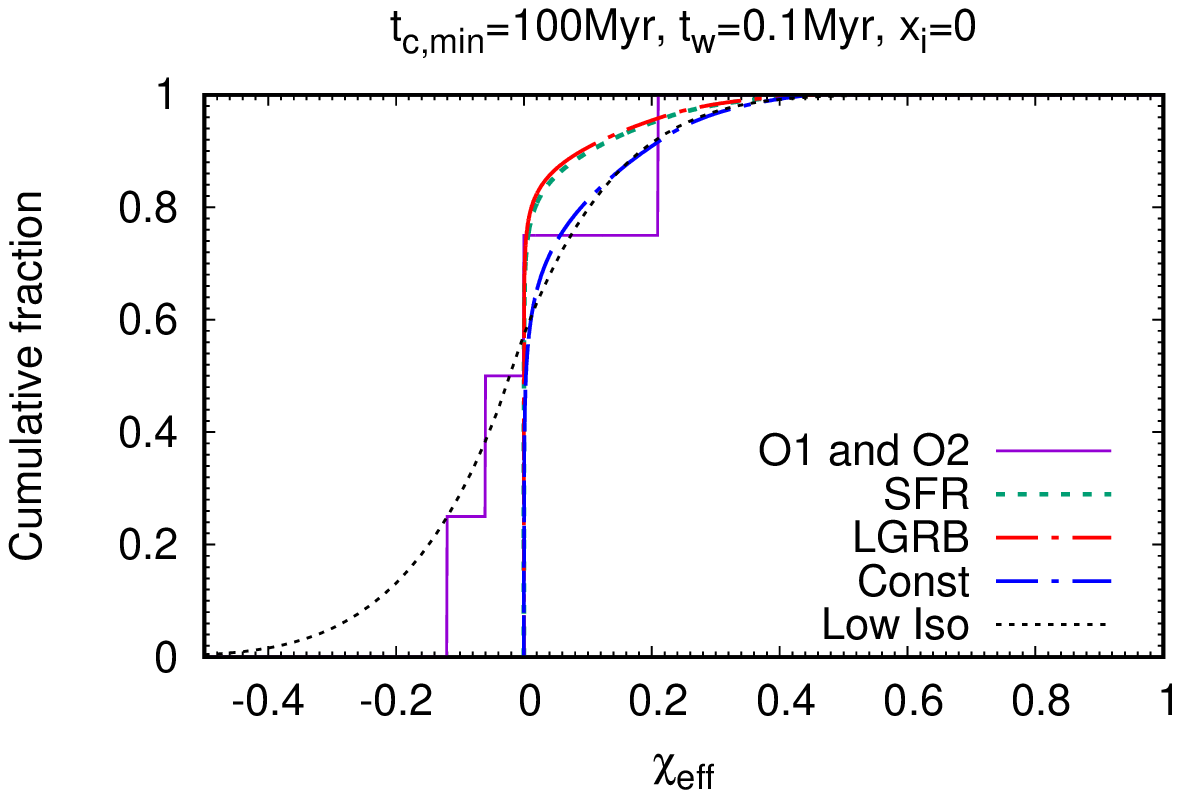}
         \includegraphics[width=6cm]{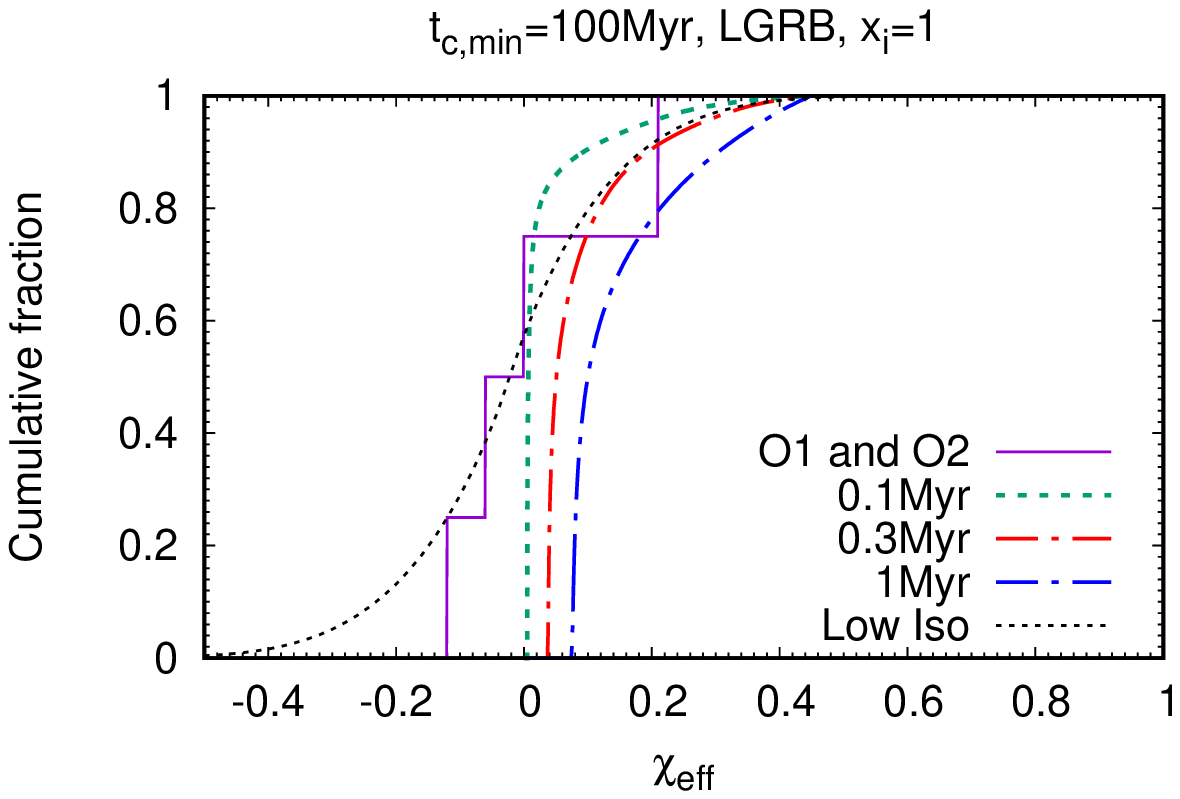}
           \caption{The cumulative $\chi_{\rm eff}$ distribution  for the O1 and O2 observing runs and for the WR binary scenario with different parameters. For the fiducial model (a blue solid line in the four upper panels), 
   the  BBH formation history  follows the cosmic SFR, the two stars are synchronized at the beginning of the WR phase,  the merger
   delay-time distribution is $\propto t^{-1}$ with a minimal time delay  of $t_{c,\,{\rm min}}=10$~Myr and the wind timescale is $t_{\rm wind}=0.3\,$Myr.  
   We set the mass ratio, $q=1$, and  $m_{\rm tot} = 60M_{\odot}$ for all the model.  Also shown as a black dashed curve is the low-isotropic spin model in \cite{farr2017arXiv}. The two bottom panels show models that deviate strongly from this fiducial choice (e.g. $t_{c,\,{\rm min}}=100$~Myr). In the left bottom panel the WR stars are not synchronized initially ($\chi_i=0$) and with the strong wind and long merger time delay distributions that follow the SFR or LGRB rates produce very narrow low $\chi_{\rm eff}$ distributions. A constant formation rate gives here a better fit to the data. In the right bottom panel the stars are initially synchronized with  a very strong wind and the long delay time the distribution is consistent with the observations. 
}
    \label{fig:KS}
  \end{center}  
\end{figure*}

\section{A comparison with observations and implications} 
\label{sec:results}

\cite{HotokezakaPiran2017ApJ} examined  the expected $\chi_{\rm eff}$ for BBH binaries of different types (see their Fig. 2). Either red or blue giant 
progenitors are easily ruled out. Even regular main sequence stars would lead to progenitors' spin values that are much larger than unity. The only possible candidates are Pop III stars, that have in fact been predicted to produce massive BBHs  \citep{kinugawa2014MNRAS} and WR stars, stripped massive stars that have lost their H and He envelopes.
 As mentioned earlier, the observed spin values are low even for those. The ``tension" appeared already in the first detection of GW150914 and it was intensified with the additional observations and in particular with the observation of GW170104.

To clarify this issue we turn now to  compare these results with predictions from a $\chi_{\rm eff}$ distribution of BBHs (see \citealt{zaldarriaga2017,HotokezakaPiran2017ApJ}). Here we focus on  WR stars.
 To form massive BHs  evolutionary scenarios require  low metallicity  progenitors  (otherwise the mass loss would be very significant). Long GRBs (LGRBs) are known to arise preferably in low metallicity hosts\footnote{From a theoretical point of view it has been suggested that
 strong winds that arise in higher metallicity progenitors will prevent fast rotation which is probably needed to produce a LGRB.}.  Therefore we use the redshift distribution of long GRB rate \citep{wanderman2010MNRAS} to estimate the rate of formation of BBH progenitors.  We also consider 
 BBH formation rates that follow the cosmic star formation rate (SFR, \citealt{madau2014ARA&A}) and a constant BBH formation rate. The resulting distribution of the SFR model is not very different from the one that follows LGRBs. On the other hand
 the results for a constant BBH formation rate are quite different as this produces a significant fraction of  binaries formed at low redshifts that have to merge rapidly and hence have small initial separations.

 We expect that the rate of  mergers 
 follows the BBH formation  rate with a time delay $t_c$ whose probability   is distributed  as $\propto t_c^{-1}$. 
We consider a minimal time delay of 1 Myr  (corresponding to an initial separation of $3 \cdot 10^{11}$cm), 10 Myr ($5.4 \cdot 10^{11}$cm), or 100 Myr ($ 10^{12}$cm) between the formation of the BBH and its merger. These differences are important as the synchronization time depends strongly on the separation and hence on $t_c$ (see Eqs. \ref{eq:Zahn} and \ref{eq:synWR}). 
We also consider different timescales of $t_{\rm wind}\equiv J_s/\dot{J}_s$, where $J_s$ is the spin angular momentum of the star,
with stronger winds corresponding to shorter $t_{\rm wind}$ values. 
 With these assumptions we obtain several probability distributions for the observed $\chi_{\rm eff}$ values.
{In general, the field binary scenario predicts a bimodal $\chi_{\rm eff}$ distribution with low and high spin peaks (see \citealt{zaldarriaga2017,HotokezakaPiran2017ApJ} for simple models and \citealt{belczynski2017,postnov2017arXiv} for population synthesis studies).
 Here the high spin peak corresponds to tidally synchronized binaries.  }
 

  Fig. \ref{fig:KS}  depicts  the integrated observed distribution of $\chi_{\rm eff}$ compared with several WR models.  One can see the large variety of the resulting $\chi_{\rm eff}$ distribution: some models  give a very large fraction of high $\chi_{\rm eff}$ mergers, while for others $\chi_{\rm eff}$ is concentrated around zero. 
  The   models with the lowest  $\chi_{\rm eff}$ distributions are those in which the progenitors (i)  are not synchronized at the beginning  of the WR phase ($\chi_i=0)$; (ii)  have a strong  wind\footnote{Note however that such winds might not be consistent with very massive remnants.} ($t_{\rm wind}=0.1$ Myr)   and (iii) have a long minimal time delay ($t_{c,{\rm min}} = 100$ Myr) - corresponding to a large initial separation.  The question whether one or two of the  progenitors is influenced by the tidal interaction is secondary as it determines the largest $\chi_{\rm eff}$  values ($> 0.4$ or $>0.8$) that have  not been observed so far. 
  
 Models with $\chi_i=1$, a  moderate wind ($t_{\rm wind} = 0.3$Myr) and a long  delay (100 Myr) in which the BBH formation rate follows the  SFR or LGRBs rates are consistent with the data (apart from the nominal negative values, of course, but those could be due to the large measurement errors). 
  The top four panels compare different models to a fiducial model in which both progenitors are spinning rapidly at the end of the common envelope phase $\chi_i=1$,  $t_{\rm wind}=0.3$ Myr, $t_{c,{\rm min}}=10 $Myr, and the BBH formation rate following the cosmic SFR. 
The lower two panels depict  more extreme models. Here we find that if all the above conditions are satisfied then the resulting $\chi_{\rm eff}$ distribution (for SFR or LGRB rate) is too narrowly centered around zero. A better fit to the data is obtained under these conditions if the BBH formation rate is a constant (see bottom left panel of Fig. \ref{fig:KS}). Even if one of the progenitor stars is synchronized at the end of the common envelope phase ($\chi_{i}=1$) a  strong enough wind ($t_{\rm wind}=0.1$ Myr) can lead to sufficient loss of angular momentum so that the final $\chi_{\rm eff}$ distributions would be very low (see bottom right panel of Fig. \ref{fig:KS}).

With just four observations it is difficult to obtain a quantitative estimate for the quality of the fit. 
Even  with the two negative nominal $\chi_{\rm eff}$ values the models that look qualitatively fine  
are  consistent with the data. 
  The KS measure, $D_{KS}=0.5$, of these models
yields chance probability  of  $\sim 20\%$, which is roughly  consistent with the $\sim15\%$ probability that $\chi_{\rm eff}$ of GW 170104 is positive
\citep{Abbott2017PRL}.  Other models that assume that even one of the stars is synchronized early on, or that $t_{\rm wind}$ is large  (0.3 or 1 Myr) or that the minimal merger time is short are qualitatively inconsistent with the observed distribution  and their quantitative chance probability $< 0.5\%$ even with this small number of events. 

The error bars of the $\chi_{\rm eff}$ estimates are not taken into account in this KS analysis.
{In order to take the relatively large measurement errors of $\chi_{\rm eff}$ into account when comparing 
different models with the data, we evaluate the odds ratios between the marginal likelihoods of different field evolution  models 
following  \cite{farr2017arXiv}.  Here we calculate the marginal likelihood of each model for the four events, $p_i(d|M)$ and then combine them as $p(d|M)=\prod_{i}p_i(d|M)$. Tables 2 and 3 list the odds ratios of these different models to the low-isotropic spin model of  \citep{farr2017arXiv}, $p(d|M)/p(d|{\rm Low\,Iso})$. {Note that this low-isotropic spin model is the most favorable one among the simple models  used in \citep{farr2017arXiv}.} None of the field binary evolution models has an odds ratio larger than unity so that the low-isotropic spin model is more consistent with the observed $\chi_{\rm eff}$ distribution than our aligned WR binary models. However, many of the models satisfying the conditions mentioned above have $p(d|M)/p(d|{\rm Low\,Iso})\gtrsim 0.1$. These cannot be ruled out with the current $\chi_{\rm eff}$ distribution of the four observed events.}

\begin{table*}
\begin{center}
\label{tab:odds1}
{\begin{tabular}{lccc}
\hline \hline
Model ($t_{c,\,{\rm min}}$) & $t_{\rm wind}=0.1$~Myr           &   $t_{\rm wind}=0.3$~Myr               & $t_{\rm wind}=1$~Myr                 \\  \hline
SFR (1Myr) & $0.12~(0.20)$ & $0.05~(0.17)$ & $0.002~(0.07)$ \\
LGRB (1Myr) & $0.13~(0.21)$ & $0.07~(0.19)$ & $0.004~(0.09)$  \\
Const (1Myr) & $0.05~(0.11)$ & $0.008~(0.06)$ & $<0.001~(0.01)$                         \\ \hline
SFR (10Myr) & $0.22~(0.33)$ & $0.10~(0.29)$ & $0.004~(0.12)$  \\
LGRB (10Myr) & $0.23~(0.33)$ & $0.13~(0.31)$ & $0.008~(0.14)$ \\
Const (10Myr) & $0.12~(0.24)$ & $0.02~(0.13)$ & $<0.001~(0.03)$ \\ \hline
SFR (100Myr) & $0.46~(0.55)$ & $0.21~(0.57)$ & $0.01~(0.26)$  \\
LGRB (100Myr) & $0.45~(0.53)$ & $0.25~(0.59)$ & $0.02~(0.31)$  \\
Const (100Myr) &$ 0.46~(0.67)$ & $0.07~(0.44)$ & $<0.001~(0.11)$  \\ 
\hline \hline 
\end{tabular}}
\end{center}
\caption{Odds ratio of the models to the low-isotropic spin model for  initially synchronized WR binaries and double (single) synchronization.}
\end{table*}

\begin{table*}
\begin{center}
\label{tab:odds2}
{\begin{tabular}{lccc}
\hline \hline
Model ($t_{c,\,{\rm min}}$) & $t_{\rm wind}=0.1$~Myr           &   $t_{\rm wind}=0.3$~Myr               & $t_{\rm wind}=1$~Myr                 \\  \hline
SFR (1Myr) & $0.12~(0.21)$ & $0.11~(0.20)$ & $0.10~(0.18)$ \\
LGRB (1Myr) & $0.13~(0.22)$ & $0.11~(0.21)$ & $0.10~(0.19)$  \\
Const (1Myr) & $0.05~(0.12)$ & $0.04~(0.10)$ & $0.03~(0.08)$                         \\ \hline
SFR (10Myr) & $0.21~(0.34)$ & $0.19~(0.32)$ & $0.18~(0.30)$  \\
LGRB (10Myr) & $0.22~(0.34)$ & $0.20~(0.33)$ & $0.19~(0.31)$ \\
Const (10Myr) & $0.13~(0.26)$ & $0.10~(0.22)$ & $0.09~(0.20)$ \\ \hline
SFR (100Myr) & $0.46~(0.56)$ & $0.41~(0.59)$ & $0.39~(0.58)$  \\
LGRB (100Myr) & $0.44~(0.54)$ & $0.40~(0.57)$ & $0.39~(0.56)$  \\
Const (100Myr) & $0.48~(0.70)$ & $0.39~(0.70)$ & $0.36~(0.66)$  \\ 
\hline \hline 
\end{tabular}}
\end{center}
\caption{Same as Table 2 but for  initially non-rotating WR binaries.}
\end{table*}

\section{Conclusions}

Before discussing the implications of these findings we turn, once more, to possible caveats. We have already argued that observations of Galactic binaries including massive BHs provide a good evidence for our model for the formation of massive BHs (no kick and no mass loss). 
{It seems that the main open issues are (i) the question of the spins of the BHs at the end of the common envelope phase, 
(ii)  the separation at the end of the common envelope phase, that determines the  tidal locking process and (iii)  the effect of  winds on the final spin. }

Turning now to the results, 
Clearly the negative observed values are inconsistent with the  model (unless there are significant kicks at the formation of the BHs). However, 
the large error bars of these measurements don't allow us to rule out any scenario. 
The observed low aligned spin values are at some ``tension" with the expectations of the standard evolutionary scenario if one takes the fiducial values we considered here  \citep[see also][]{kushnir2016MNRAS,HotokezakaPiran2017ApJ,zaldarriaga2017}. However, even for these parameters the small number statistics is insufficient  to make any clear conclusions. More important  are the uncertainties in the outcome of the common envelope phase (whether the stars are synchronized or not at the end of this phase), in the strength of the wind at the late stages of the evolution of these massive stars, in the minimal time delay for mergers (corresponding to the minimal separation and hence to the importance of the tidal locking process) and finally in the tidal synchronization process itself. 
For example, it is clear that strong enough winds would reduce the final spin of the stars. However, one may wonder if such strong winds are consistent with the very massive BHs observed. 
Note that $t_{\rm wind}=0.1$ Myr roughly corresponds to a mass loss of  $10^{-4.5} m_\odot$/yr  which is at the level of the strong winds of the observed WR stars~\citep{crowther2007ARA&A,vink2017}. 


Both the SFR and  LGRB rates are favorable as proxies for the BBH formation rate. In both cases most of the formation takes place at early times, allowing for large initial separations. The resulting $\chi_{\rm eff}$ distributions arising from these two scenarios are practically indistinguishable. A constant BBH formation rate implies more recent formation events and hence shorter merger times leading to  larger   $\chi_{\rm eff}$ values. Still with 
extreme parameters even this distribution can be made consistent with the current data. 

 High redshift WR stars are the best candidates for being progenitors with low $\chi_{\rm eff}$. They gain from having a long merger times, that allows them to begin with a relatively large separation that implies much weaker synchronization.   
 However, this is not enough and strong or moderate winds (for  progenitors that are non-rotating at the end of the common envelope phase)  are essential  for consistency with the current distribution. A longer  minimal time delay (corresponding to larger separations at the birth of  BBHs) helps, but is insufficient to lead to consistency. 
  

To conclude  we note that a comparison of the currently observed O1 and O2 $\chi_{\rm eff}$ values with the models show some tension, however it does not rule out evolutionary models based on WR stars.  In fact some  models are almost as consistent as the best fitted low-isotropic spin model of \cite{farr2017arXiv}. While many models predict a significant fraction ($>25\%$) of large ($>0.4$ for singly synchronized and  $>0.8$ for double synchronization) $\chi_{\rm eff}$ events, some  produce distributions that are concentrated around positive very low $\chi_{\rm eff} $ values.  The question which models are consistent depends on largely unexplored late stage evolution of very massive stars. 
Given these results it seems that while a significant fraction of high $\chi_{\rm eff}$ mergers will strongly support the field evolutionary scenario, lack of those will be hard to interpret. It may indicate another scenario or, for example,   strong winds that remove the spin angular momentum. On the other hand a significant fraction of negative $\chi_{\rm eff} $ merger will be difficult to reconcile with this scenario, unless the BHs' angular momentum is dominated by very strong natal kicks. 
  

\section*{Acknowledgments}

KH is supported by the Flatiron Fellowship at the Simons Foundation.
TP is supported by an advanced 
ERC grant TReX and by the ISF-CHE I-Core center of excellence for research in Astrophysics. 


\end{document}